\documentclass[12pt]{article}

\usepackage{setspace}
\usepackage{epsfig}
\usepackage{cite}
\usepackage{amssymb}
\begin{document}
\title{Statistical thermodynamics of adhesion points in supported
membranes} \author{Oded Farago\\ Department of Biomedical Engineering
and Ilse Katz Institute for Nanoscale\\ Science and Technology,
Ben-Gurion University of the Negev,\\ Be'er Sheva 84105, Israel.}
\maketitle
\begin{abstract}

Supported lipid membranes are useful and important model systems for
studying cell membrane properties and membrane mediated processes. One
attractive application of supported membranes is the design of phantom
cells exhibiting well defined adhesive properties and receptor
densities. Adhesion of membranes may be achieved by specific and
non-specific interactions, and typically requires the clustering of
many adhesion bonds into ``adhesion domains''. One potential mediator
of the early stages of the aggregation process is the Casimir-type
forces between adhesion sites induced by the membrane thermal
fluctuations. In this review, I will present a theoretical analysis of
fluctuation induced aggregation of adhesion sites in supported
membranes. I will first discuss the influence of a single attachment
point on the spectrum of membrane thermal fluctuations, from which the
free energy cost of the attachment point will be deduced. I will then
analyze the problem of a supported membrane with two adhesion
points. Using scaling arguments and Monte Carlo simulations, I will
demonstrate that two adhesion points attract each other via an
infinitely long range effective potential that grows logarithmically
with the pair distance. Finally, I will discuss the many-body nature
of the fluctuation induced interactions. I will show that while these
interactions alone are not sufficient to allow the formation of
aggregation clusters, they greatly reduce the strength of the residual
interactions required to facilitate cluster formation. Specifically,
for adhesion molecules interacting via a short range attractive
potential, the strength of the direct interactions required for
aggregation is reduced by about a factor of two to below the thermal
energy $k_BT$.

\end{abstract}
\vspace{0.45cm}

\newpage

\section{Introduction}
\label{sec:intro}

Fatty acids and other lipids are essential to every living
organism. Because of their amphiphilic nature, they spontaneously
self-assemble into bilayer membranes that define the limits of cells
and serve as permeability barrier to prevent proteins, ions and
metabolites from leaking out of the cell and unwanted toxins leaking
in \cite{Lipowsky_Sackmann}. In euokaryotic cells, membranes also
surround the organelles allowing for organization of biological
processes through compartmentalization. In addition, biological
membranes host numerous proteins that are crucial for the mechanical
stability of the cell, and which carry out a variety of functions such
as energy and signal transduction, communication, and cellular
homeostasis \cite{Alberts}.

An important aspect of biological membranes is that they are typically
not free but rather confined by other surrounding membranes, adhere to
other membranes, and attach to elastic networks like the cytoskeleton
and the extracellular matrix. Several model systems with reduced
compositional complexity have been designed to mimic biological
membranes. These biomimetic systems include phospholipid bilayers
deposited onto solid substrates (solid-supported
membranes)~\cite{Saldit:2005}, or on ultra-thin polymer supports
(polymer-supported membranes)~\cite{Tanaka_Sackmann:2005}.  Placing a
membrane on a flat substrate allows for the application of several
different surface sensitive techniques, including atomic force   
microscopy, x-ray and neutron diffraction, ellipsometry, nuclear
magnetic resonance, and others \cite{rinia:2006}.  With the aid of
biochemical tools and generic engineering, supported membranes can be
functionalized with various membrane-associated
proteins~\cite{Tanaka_Sackmann:2006}. Synthetic supported membranes
with reconstituted proteins are increasingly used as controlled
idealized models for studying key properties of cellular
membranes~\cite{Girard:2007}. They provide a natural environment for  
the immobilization of proteins under nondenaturating conditions and in
well-defined orientations~\cite{Salafsky:1996a}. Another attractive
application of supported membranes is the design of phantom cells
exhibiting well defined adhesive properties and receptor
densities~\cite{sackmann:1996}. Using advanced imaging techniques,
detailed information can be obtained about the structure of the
adhesion zone between the receptor-functionalized supported membrane
and ligand-containing vesicles that can bind to the supported membrane
\cite{kloboucek:1999,kaizuka:2004}. These studies provide insight into
the dynamics of adhesion processes and the molecular interactions
involved in cell adhesion
\cite{smith:2006,ananth:2007}. Understanding these interactions is   
crucial for the development of drug delivery systems that depend on 
efficient adhesion between a liposome and the plasma membrane of the
target cell.

Adhesion between two membranes or between a membrane and another
surface can, in principle, be facilitated by non-specific attractive
interactions (e.g., Coulomb and van der Waals interactions)
\cite{swain:1999,komura:2003,mecke:2003,hoopes:2008}. Cell adhesion,
however, is usually caused by highly specific receptor molecules
located at the outside of the plasma membrane of the cell, that can
bind to specific ligands on the opposite surface
\cite{beckerle:2001,lauff_inderman:1995}.  Typically, the area density
of the receptor molecules located at the outside of the plasma
membrane is rather low which does not lead to efficient adhesion.
However, when facing a surface with enough ligands, the receptors may
cluster into highly concentrated adhesion domains to establish much
stronger binding \cite{smith:2007,weikl:2009}. Formation of adhesion
clusters occurs in many biological processes
\cite{lenne_nicolas:2009}, including the binding of white blood cells
to pathogens \cite{naggli:2003}, cadherin-mediated adhesion of
neighboring cells \cite{giehl_menke:2008}, and focal adhesion of cells
to the extracellular matrix \cite{geiger:2001}. Many biophysical
aspects of specific adhesion processes, ranging from the cooperativity
in adhesion cluster formation to the influence of stochastic processes
such as the ligand-receptor reaction kinetics, have been and continue
to be studied theoretically using various models
\cite{moreira:2003,zhang:2003,gov:2004,gruhn:2005,longo:2005,moore:2006,lin:2006,krobath:2007,merath:2007,lamblet:2008,reister:2008,spatz:2008}.

\begin{figure}[h]
  {\centering
  \hspace{1cm}\epsfig{file=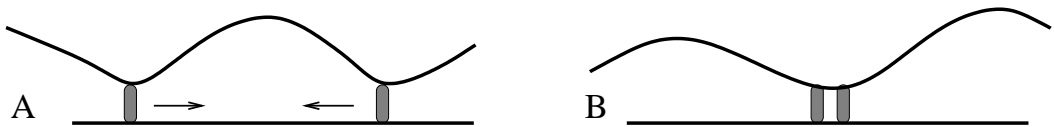,width=14cm}}
\caption{(A) Schematic of a membrane attached by two distant adhesion
bonds to an underlying surface. There is an entropy penalty associated
with each adhesion bonds due to the restrictions imposed on the
membrane thermal fluctuations in their vicinity. (B) The entropy cost
can be minimized by bringing the adhesion bonds close to each other,
in which case the thermal fluctuations become limited at only one
location. The increase in the entropy in (B) compared to (A) is the
origin of the attractive fluctuation-induced interactions between the
adhesion bonds.}
\label{fig:1}
\end{figure}

Adhesion induced domain formation requires some attractive
intermolecular interactions between the receptor-ligand pairs. These
interactions include both {\em direct}\/ and {\em membrane-mediated}\/
contributions. The former are typically described by pairwise
potentials which are infinitely repulsive at very small molecular
separations and attractive at somewhat longer (but still finite)
distances \cite{israelachvili}. Their effect can, therefore, be
studied in the framework of the thoroughly researched lattice-gas
model \cite{simon}.  In contrast, much less is known about the
membrane-mediated mechanism, which has been proposed by Braun {\em et
al.}\/ to explain to formation of gap junctional plaque at cell-cell
interfaces \cite{boa}, and whose origin can be understood as follows:
Consider two adhesion bonds between two membranes or between a
membrane and a surface (Fig.~\ref{fig:1}(A)). The adhesion points
restrict the thermal height fluctuations of the membrane in their
vicinity. This entropy loss can be minimized if the two adhesion bonds
are brought to the same place (Fig.~\ref{fig:1}(B)), in which case the
membrane becomes pinned at only one place rather than two. The
membrane fluctuations, thus, induce an attractive potential of mean
force between the adhesion bonds. This effect is often named after
Casimir who predicted the existence of an attractive force between two
conducting plates, due to quantum fluctuations of the electromagnetic
field in the intervening space \cite{casimir}. Later, Fisher and de
Gennes generalized this concept to classical interactions induced by
thermal fluctuations in soft matter systems
\cite{fisher_degennes}. For bilayer membranes, there is a great body
of theoretical work on the Casimir effect between trans-membrane
proteins (see review in \cite{kardar:1999}, and refs.~therein). Just
like adhesion bonds, membrane inclusions represent a ``constraint'' on
the shape of the membrane and, therefore, one should expects that they
also interact with each other through Casimir-like interactions. In
addition to the fluctuation-induced forces, the inclusions also
experience other membrane-mediated interactions which arise from the
membrane curvature elasticity and from the packing of the lipids near
the inclusions' surfaces (see review in \cite{deserno_review}, and
refs.~therein). These other types of membrane-mediated interactions
are also expected to exist between membrane adhesion bonds.

\begin{figure}[h]
  {\centering
  \hspace{1cm}\epsfig{file=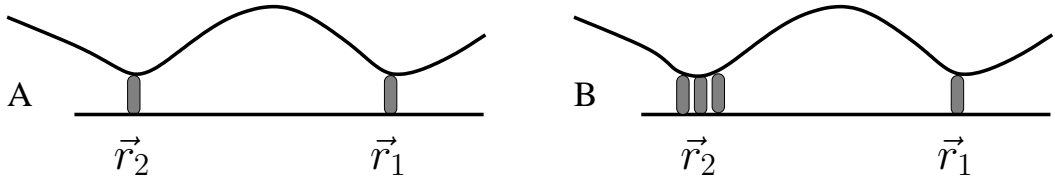,width=14cm}}
\caption{(A) Schematic of a supported membrane with two adhesion
located at $\vec{r}_1$ and $\vec{r}_2$. (B) Similar to (A), but with a
three-bond cluster instead of a single adhesion bond in
$\vec{r}_2$. The adhesion bond in $\vec{r}_1$ is equally attracted (by
a Casimir-like force) to the adhesion bond located in $\vec{r}_2$ in
(A) and to the cluster of three adhesion bonds shown in (B).}
\label{fig:2}
\end{figure}

The fundamental difficulty in attempting to provide a
statistical-mechanical analysis of the aggregation behavior of the
adhesion bonds is the need to integrate out the membrane degrees of
freedom and write down the potential of mean force as a function of
the coordinates of the adhesion sites
$\phi(\vec{r}_1,\vec{r}_2,\vec{r}_3,\ldots,\vec{r}_N)$. This is a
non-trivial problem since the membrane-mediated potential
$\phi(\vec{r}_1,\vec{r}_2,\vec{r}_3,\ldots,\vec{r}_N)$ is a many-body
potential which cannot be expressed as the sum of two body terms. The
many-body nature of
$\phi(\vec{r}_1,\vec{r}_2,\vec{r}_3,\ldots,\vec{r}_N)$ is best
illustrated by the following example: Consider the configuration shown
in Fig.~\ref{fig:2}(A) with two adhesion bonds at located at
$\vec{r}_1$ and $\vec{r}_2$ and, in comparison, the one shown in
Fig.~\ref{fig:2}(B) with a single bond at $\vec{r}_1$ and a cluster of
three bonds around $\vec{r}_2$. Clearly, the spectrum of membrane
thermal fluctuations in both cases is quite the same and, therefore,
the adhesion bond located at $\vec{r}_1$ is attracted to the
three-point cluster in \ref{fig:2}(B) by the same force to which it is
attracted to the single adhesion point in \ref{fig:2}(A). If
$\phi(\vec{r}_1,\vec{r}_2,\vec{r}_3,\ldots,\vec{r}_N)$ was the sum of
pair interactions, the force in Fig.~\ref{fig:2}(B) would be three
times larger than the force in \ref{fig:2}(A).

\section{Lattice-gas model for adhesion bonds in supported membranes}
\label{sec:lattice-gas}

What is the difference between the aggregation of adhesion bonds in
supported membranes and the traditional process of gas to liquid
condensation? Condensation phase transitions are usually associated
with a competition between the mixing entropy $S$ which is higher in
the dilute gas phase, and the interaction energy $U$ which is lower in
the condensed liquid state. The equilibrium phase corresponds to the
minimum of the free energy $F=U-TS$, where $T$ is the temperature of
the system. At high $T$, the free energy $F$ is ``entropy-dominated''
and equilibrium is attained in the gas phase. Conversely, at low $T$,
the free energy is ``energy-dominated'' and, therefore, the condensed
phase becomes thermodynamically more favorable. The liquid-gas phase
transition can be analyzed in the framework of an Ising-like model of
identical particles that populate a lattice. Excluded volume
interactions between the particles are represented by the fact each
lattice site can be occupied by no more than one particle. When two
particles occupy nearest-neighbor sites, they interact in a pairwise
fashion with an attractive energy $-\epsilon$. Denoting the occupancy
of a lattice site by $s_i$, with $s_i=0$ for an empty site and $s_i=1$
for an occupied site, the Hamiltonian of the lattice-gas model is
given by
\begin{equation}
{\cal H}_{\rm LG}=-\epsilon\sum_{i,j}s_is_j,
\label{eq:latticegas}
\end{equation}
where the sum runs over all the pairs of lattice nearest neighbor
sites. The phase diagram of the lattice-gas model is well known. There
exists a critical value $\alpha_c$ such that if the interaction energy
$\epsilon<\alpha_c k_BT$, the particles will be distributed uniformly
within the lattice. Above this critical value, $\epsilon>\alpha_c
k_BT$, a uniform distribution of the particles is observed only at low
concentrations of particles (``gas phase''), but upon increasing the
concentration of particles, the system undergoes a first order phase
transition and a second coexisting phase appears with a considerably
larger concentration (``condensed phase'').

As discussed in the previous section, the aggregation process of
adhesion domains involves an additional attractive potential of mean
force resulting from the membrane thermal fluctuations. A lattice
model where each lattice particle represents an adhesion bond in a
supported membrane can, therefore, be introduced by supplementing
Eq.(\ref{eq:latticegas}) with an energy term corresponding to the
fluctuation-induced interactions. Since the functional form of this
many-body potential of is yet unknown, we would, at this moment,
introduce it via a general potential function $\phi$ that depends on
the coordinates of the lattice particles:
\begin{equation}
{\cal
H}=-\epsilon\sum_{i,j}s_is_j+\phi\left(\left\{s_i\right\}\right).
\label{eq:latticegas2}
\end{equation} 
Our first task must be to derive an expression for
$\phi\left(\left\{s_i\right\}\right)$. Once this is accomplished, one
can attempt to analyze the statistical mechanical properties of the
model and address the question appearing at the beginning of section
\ref{sec:lattice-gas}. One particular issue that we would like to
address is whether the fluctuation induced attractive potential (which
is of entropic origin) can win the competition against the repulsive
force originating from the mixing entropy?  In other words, can
adhesion clusters form for purely entropic grounds, i.e. for
$\epsilon=0$ in Eq.(\ref{eq:latticegas2})? Gas to liquid condensation
transitions are generally believed to involve energy vs.~entropy
competition \cite{frenkel}, but purely entropic phase transitions from
a fluid (disordered) phase into a crystalline (ordered) phase are
known to exist. Hard sphere systems, for instance, undergo a first
order phase transition from a low density fluid phase into a high
density solid phase \cite{phys_today}. This transition results from
the competition between two entropies - the configurational mixing
entropy which is higher in the disordered phase, and the entropy
associated with the free volume available for each sphere, which is
higher in the ordered crystal.

\section{Statistical mechanics of a membrane with one adhesion point}
\label{sec:onepoint}

\begin{figure}[h]
  {\centering
  \hspace{1cm}\epsfig{file=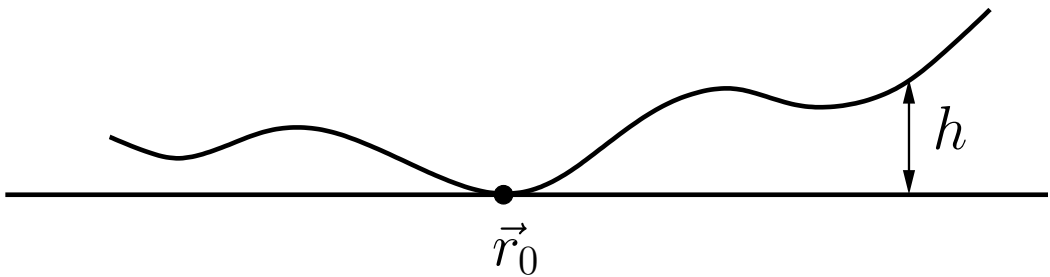,width=14cm}}
\caption{(A) Schematic picture of the model system consisting of a
membrane that fluctuates above a flat impenetrable surface to which it
is pinned at a single point.}
\label{fig:3}
\end{figure}

We start our analysis by considering the system shown schematically in
Fig.~\ref{fig:3}, consisting of a membrane with bending rigidity
$\kappa$ that fluctuates above a flat impenetrable surface
\cite{farago1}. Let $h\left(\vec{r}\right)\geq 0$ be the height of the
membrane with respect to the surface, and assume that the membrane is
pinned to the surface at one fixed point located at $\vec{r}_0$
($h\left(\vec{r}_0\right)=0$). The elastic curvature energy of the
membrane is given by the Helfrich effective Hamiltonian
\cite{helfrich:1973}
\begin{equation}   
{\cal H}_{\rm hel} = \int\,
\left[\frac{\kappa}{2}\left(\nabla^2 h \right)^2\right]\Phi\left(h\right)
\delta\left[h\left(\vec{r_0}\right)\right]d^2\vec{r},
\label{eq:helfrich}
\end{equation}
where $\Phi$ represents the hard wall constraint due to the surface
($\Phi=1$ for $h\geq 0$, and $\Phi=+\infty$ for $h<0$), $\delta$ is
the Dirac delta-function, and the integration is taken over the cross
sectional (projected) area of the membranes of size $L^2$. 

\begin{figure}[t]
  {\centering
  \hspace{0cm}\epsfig{file=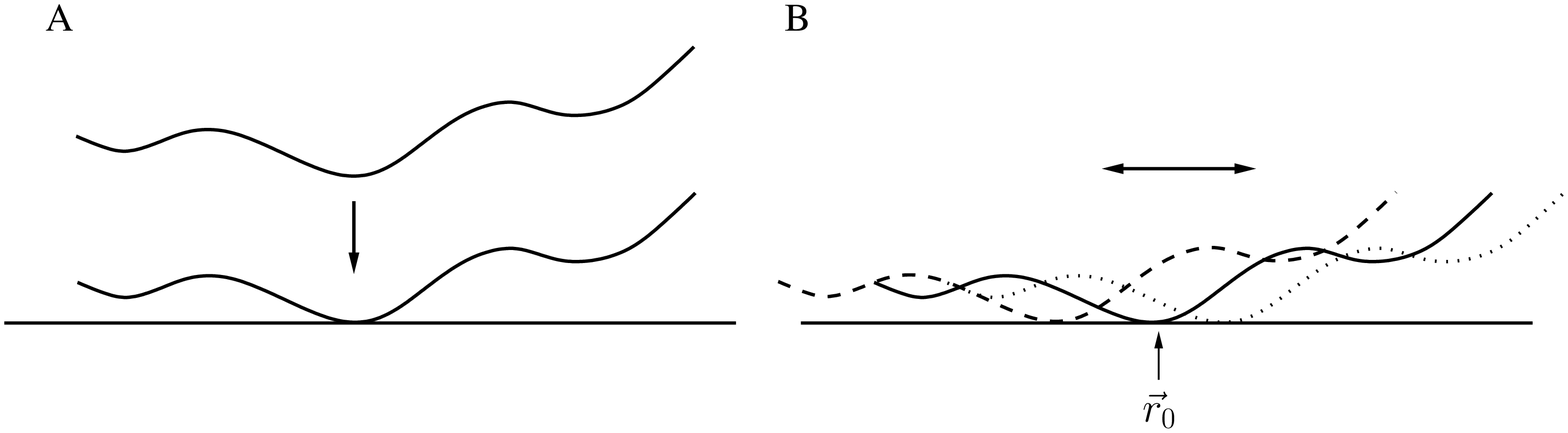,width=16cm}}
\caption{(A) A freely fluctuating membrane can be always translated
vertically such that the point at which its height function
$h\left(\vec{r}\right)$ attains its global minimum is on a flat
surface and the rest of the membrane is above the surface. (B) A
freely fluctuating membrane can be also translated horizontally. All
the membrane configurations generated in this way are similar to each
other, and the one for which the point of absolute minimum is at
$\vec{r}_0$ (represented by the solid line) is identical to the pinned
membrane configuration shown in {\protect Fig.~\ref{fig:3}}.}
\label{fig:4}
\end{figure}

One can calculate the partition function $Z$ corresponding to
Hamiltonian (\ref{eq:helfrich}), by considering the Helfrich effective
Hamiltonian of a {\em freely fluctuating}\/ membrane
\begin{equation}   
{\cal H}^{0}_{\rm hel} = \int\,
\left[\frac{\kappa}{2}\left(\nabla^2 h \right)^2\right]d^2\vec{r}. 
\label{eq:helfrich2}
\end{equation}
In this case, the associated partition function $Z_{\rm free}$ is
readily calculated by introducing the Fourier transformation of
$h\left(\vec{r}\right)$: $h_q=(1/L^2)\int
h\left(\vec{r}\right)\exp(i\vec{q}\cdot\vec{r})$, which decompose
Hamiltonian (\ref{eq:helfrich2}) into the sum of independent harmonic
oscillators
\begin{equation}   
{\cal H}^{0}_{\rm hel} = \frac{l^4}{L^2}\sum _{\vec{q}}
\frac{\kappa}{2}q^4|h_q|^2,
\label{eq:helfrich3}
\end{equation}
where $l$ is a microscopic length scale of the order of the bilayer
thickness. Hamiltonian (\ref{eq:helfrich}) which also includes the
functions $\Phi$ and $\delta$ cannot be diagonalized in the same
manner. However, one can relate the partition function $Z$ of
Hamiltonian (\ref{eq:helfrich}) with the partition $Z_{\rm free}$ of
the free membrane Hamiltonian (\ref{eq:helfrich2}), by using the
following simple argument. The energy of a freely fluctuating membrane
is invariant with respect to rigid-body transformations such as a
vertical translation ($h\left(\vec{r}\right)\rightarrow
h\left(\vec{r}\right)-h_0$) of the membrane's center of
mass. Therefore, one can draw a flat surface and translate the free
membrane such that the global minimum of its height function coincides
with the surface (see Fig.~\ref{fig:4}(A)). The vertically translated
free membrane looks very similar to the pinned membrane shown in
Fig.~\ref{fig:3}(A). The only difference between them is that the
former can also glide over the surface (see Fig.~\ref{fig:4}(B)),
while the latter is pinned at a fixed position on the surface. This
suggests that the pinning point effectively eliminates the membrane
horizontal translational degree of freedom. In a statistical
mechanical language, the configurational phase space of the pinned
membranes is smaller than, yet {\em similar}\/ to, the phase space of
a free membrane. Each subspace of identical free membrane
configurations, like the ones shown in Fig.~\ref{fig:4}(B), includes
one pinned membrane configuration - the configuration where the
minimum of $h\left(\vec{r}\right)$ is at the pinning point $\vec{r}_0$
(or, more precisely, within a microscopic area of size $l^2$ around
the pinning point, where $l$ is the spatial resolution of the
continuum model). This pinned membrane configuration occupies a
fraction $(l/L)^2$ of the corresponding larger free membrane
configurational subspace, which implies that the partition functions
of the two systems are related by $Z=(l/L)^2 Z_{\rm free}$. The free
energy is obtained from
\begin{equation}
F=-k_BT\ln\left(Z\right)=-k_BT\ln\left(Z_{\rm free}\right)
+2k_BT\ln\left(\frac{L}{l}\right)
\end{equation}
The first term on the right hand side is the free energy of the free
membrane whose elastic energy is given by Helfrich Hamiltonian
(\ref{eq:helfrich}). The second term,
\begin{equation}
F_{\rm attachment,1}=2k_BT\ln\left(\frac{L}{l}\right),
\label{eq:attachmentfenergy}
\end{equation}
is the free energy cost of attaching the membrane to the surface at
one point.

\begin{figure}[t]
  {\centering
  \hspace{2.5cm}\epsfig{file=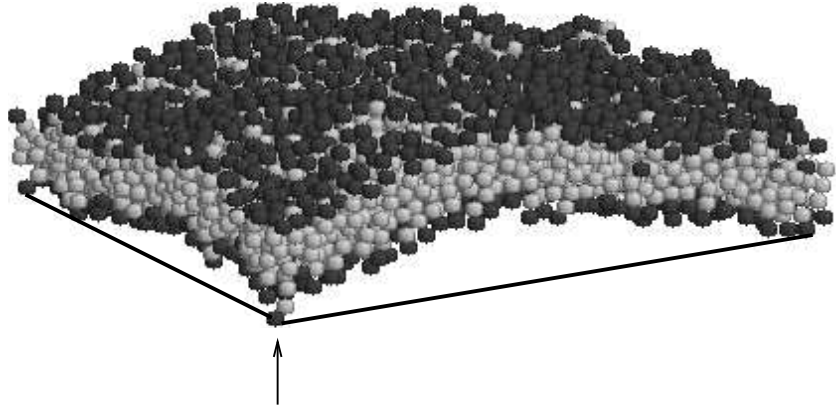,width=11cm}}
\caption{Equilibrium configuration of a membrane consisting of 2000
lipids. Each lipid is represented by a trimer of one ``hydrophilic''
bead (dark gray sphere) and two ``hydrophobic'' beads (light gray
spheres). The membrane is fluctuating above a plane surface (frame
indicated by a thick black line), while one of the hydrophilic beads
(the black sphere appearing at the front of the figure and indicated
by an arrow) is held on the surface at a fixed position.}
\label{fig:5}
\end{figure}

\begin{figure}[t]
  {\centering
  \hspace{1cm}\epsfig{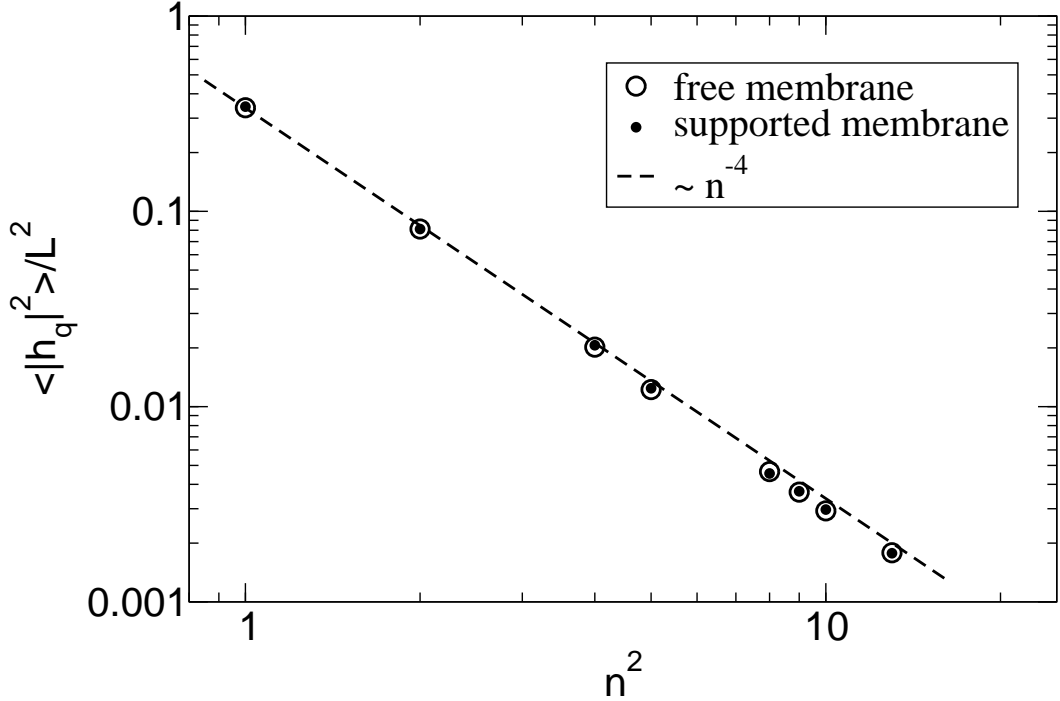}}
\caption{The mean square amplitude of the thermal height fluctuations
as a function of the wavenumber $n$. The results from the supported
membrane simulations are shown by small solid circles. These results
are essentially identical to those obtained from simulations of a free
membrane which are represented by larger open circles. The dashed line
indicates the asymptotic $\langle|h_q|^2\rangle\sim n^{-4}$ power law
for small $n$. (adapted from \cite{farago1})}
\label{fig:6}
\end{figure}

Following the above argument leads to a very interesting
conclusion. Because of the similarity mapping that exists between the
configurational phase spaces of the two problems, the statistical
properties of the pinned and the free membrane must be identical to
each other. This surprising result can be demonstrated by using an
implicit-solvent coarse-grained (ISCG) bilayer model which enables
molecular simulations of mesoscopically large bilayer membranes over
relatively large time-scales \cite{is_1,is_2,is_3}.  Towards this end,
we ran two independent Monte Carlo (MC) simulations - one of a free
membrane (without a surface) and one of a membrane supported by a flat
impenetrable surface. A snapshot from the supported membrane
simulations is shown in Fig.~\ref{fig:5}. Each lipid molecule is
represented in the model by a short string of three spherical beads,
where one of the beads (depicted as a dark gray sphere in
Fig.~\ref{fig:5}) represents the hydrophilic head group and two beads
(light gray spheres in Fig.~\ref{fig:5}) represent the hydrophobic
tail of the lipid. In the supported membrane simulations, the head
bead of one of the lipids (appearing in the corner at the front of the
figure and indicated by an arrow) was fixed to a flat surface which
the lipids were not allowed to cross.  We measured the Fourier
spectrum of the membrane height function. For the free membrane, the
application of the equipartition theorem to the Fourier-space
representation of the Helfrich Hamiltonian (\ref{eq:helfrich3}) yields
the following relationship between the mean squared amplitude of the
Fourier modes (``spectral intensity'') and the wave-vector $q$:
\begin{equation}
\left\langle|h_q|^2\right\rangle=\frac{k_BTL^2}{\kappa l^4q^4}.
\label{eq:eqpartition}
\end{equation}
Fig.~\ref{fig:6} depicts the results of our MC simulations for the
spectral intensity vs.~the wavenumber $n=qL/(2\pi)$. The figure show
that, in agreement with our predictions: (i) the free (open circles)
and pinned (solid circles) membranes exhibit the same statistics of
thermal height fluctuations, and (ii) the spectral intensities of both
membranes follow the $n^{-4}$ power-law dependence anticipated by
Eq.(\ref{eq:eqpartition}) (dashed line).

\begin{figure}[h!]
  {\centering
  \hspace{2cm}\epsfig{file=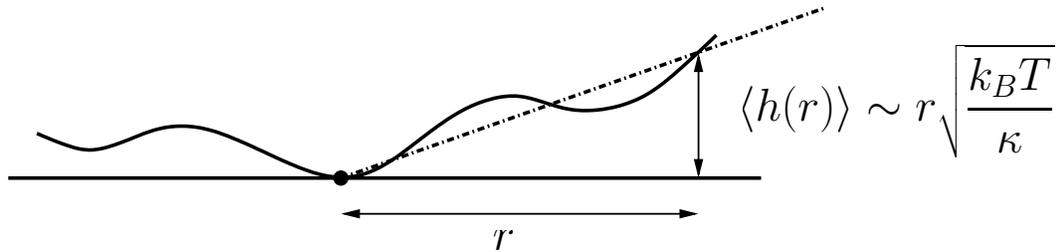,width=14cm}}
\caption{The fact that the statistics of thermal height fluctuations
is not affected by the single pinning point implies that the typical
height of the fluctuations scales linearly with the distance from the
pinning site.}
\label{fig:7}
\end{figure}

One can reverse the argument and derive
Eq.(\ref{eq:attachmentfenergy}) starting from the assumption that the
spectral intensity of the supported membrane is identical to that of a
free membrane and, therefore, can be described by
Eq.(\ref{eq:eqpartition}). The derivation proceeds as follows: First,
from Eq.(\ref{eq:eqpartition}), one can straightforwardly show that
the typical height at which the membrane undulates above the surface
at a distance $r$ away from the pinning point scales linearly with $r$
\cite{farago1,bruinsma:1994} (see Fig.~\ref{fig:7}):
\begin{equation}
u(r)\equiv\left\langle h\left(r\right)\right\rangle\sim
r\sqrt{\frac{k_BT}{\kappa}}.
\label{eq:linear}
\end{equation} 
There is a repulsive force acting between the fluctuating membrane and
the underlying surface, caused by their mutual steric
hindrance. Helfrich \cite{helfrich:1978} showed that the associated
repulsive interaction free energy density (per unit area) has the
following scaling behavior $V(r)\sim (k_BT)^2/\kappa
u(r)^2$ which, together with Eq.(\ref{eq:linear}), yields
\begin{equation}
V(r)\sim \frac{k_BT}{r^2}.
\label{eq:fenergydensity}
\end{equation}
By integrating this energy density over the projected area of the
membrane, one derives Eq.(\ref{eq:attachmentfenergy}) up to a
numerical prefactor
\begin{equation} 
F_{\rm attachment,1}= \int\, V(r)d^2\vec{r}
\sim \int_l^L 2\pi r\frac{k_BT}{r^2}dr
=Ck_BT \ln\left(\frac{L}{l}\right).
\label{eq:attachmentfenergy2}
\end{equation}
To set $C=2$, as in Eq.(\ref{eq:attachmentfenergy}), one needs to
replace the scaling relation Eq.(\ref{eq:fenergydensity}) with the
equality
\begin{equation}
V(r)=\frac{1}{\pi}\frac{k_BT}{r^2}.
\label{eq:fenergydensity2}
\end{equation}

\section{Fluctuation induced attraction between two adhesion points}
\label{sec:twopoints}

As noted by Helfrich \cite{helfrich:1978}, the free energy density
Eq.(\ref{eq:fenergydensity2}) due to the steric hindrance between the
two surfaces (i.e., the fluctuating membrane and the underlying
supporting surface) is directly related to the rate of collisions
between them. In other words, the probability density that the
membrane hits the supporting surface at a distance $r$ from the
pinning point exhibit the same dependence on $r$ as $V(r)$:
\begin{equation}
p\left[h\left(\vec{r}\right)=0\right]\sim \frac{1}{r^2}.
\label{eq:prob1}
\end{equation}
This relationship provides the information needed for calculating the
fluctuation induced attractive potential between two adhesion points.
This is done by regarding the point of collision between the membrane
and the surface as a second pinning point which can diffuse across the
surface. In this context, the probability density
$p\left[h\left(\vec{r}\right)=0\right]$ is identified with the pair
correlation function between the adhesion points which, therefore,
also follows the scaling form
\begin{equation}
g\left(\vec{r}\right)\sim \frac{1}{r^2}.
\label{eq:pair1}
\end{equation}
By definition, the pair potential of mean force is given by
\begin{equation}
\phi\left(\vec{r}\right)\equiv-k_BT\ln\left[g\left(\vec{r}\right)\right]
=2k_BT\ln\left(r\right),
\label{eq:pairpotential}
\end{equation}
which is an infinitely long range attractive potential that does not
depend of the bending rigidity of the membrane, $\kappa$.

\begin{figure}[t]
  {\centering
  \hspace{1cm}\epsfig{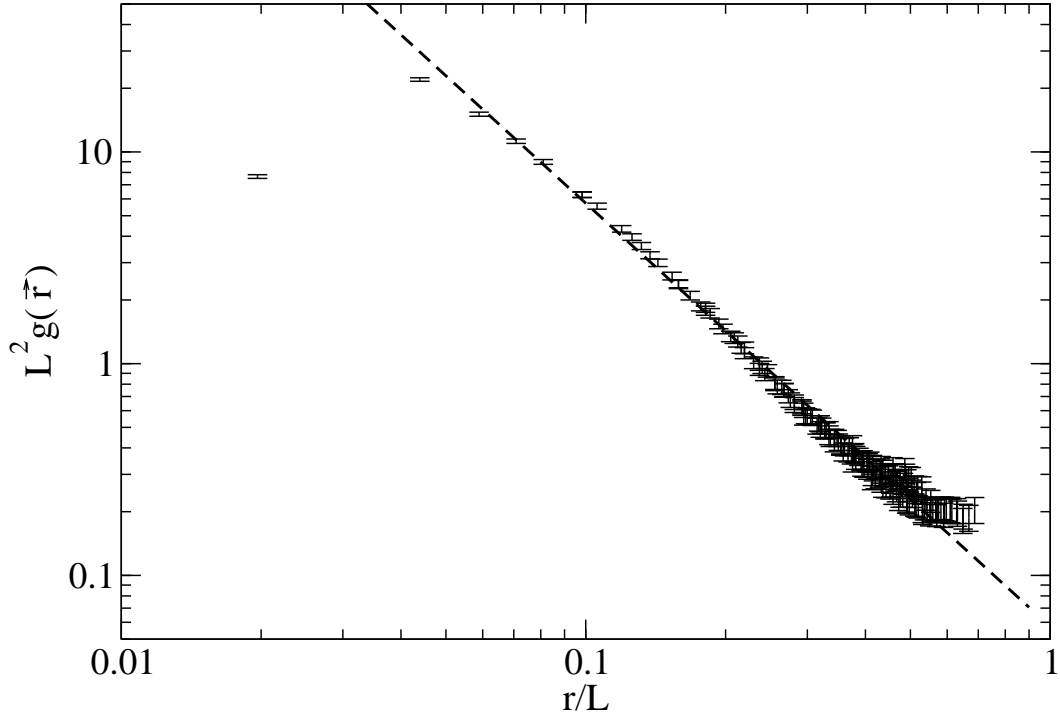}}
\caption{The pair correlation function, $g\left(\vec{r}\right)$, of a
non-stressed membrane vs.~the pair distance $r$. The slope of the
dashed straight line is $-2$. (adapted from {\protect \cite{farago2}})}
\label{fig:8}
\end{figure}

The validity of Eq.(\ref{eq:pair1}) can be tested by using MC
simulations of the ISCG model shown in Fig.~\ref{fig:5} with two lipid
heads attached to surface - one fixed at the origin and the other
allowed to diffuse on the flat surface. The pair correlation function
is then directly computed by sampling the position of the mobile
adhesion point. Our results \cite{farago2}, which are shown in
Fig.~\ref{fig:8}, agree very well with Eq.(\ref{eq:pair1}). The slope
of the straight line on the log-log plot is equal to $-2$. The
deviations from the power law behavior $g(\vec{r})\sim 1/r^2$ at small
values of $r$ ($r/L<0.05$) are related to the breakdown of the
continuum description of the Helfrich Hamiltonian at small spatial
scales. At small separations, the molecular nature of the lipids
becomes important and the radial pair distribution function is
dominated by the depletion shells around the lipids.

\begin{figure}[t]
  {\centering
  \hspace{1cm}\epsfig{file=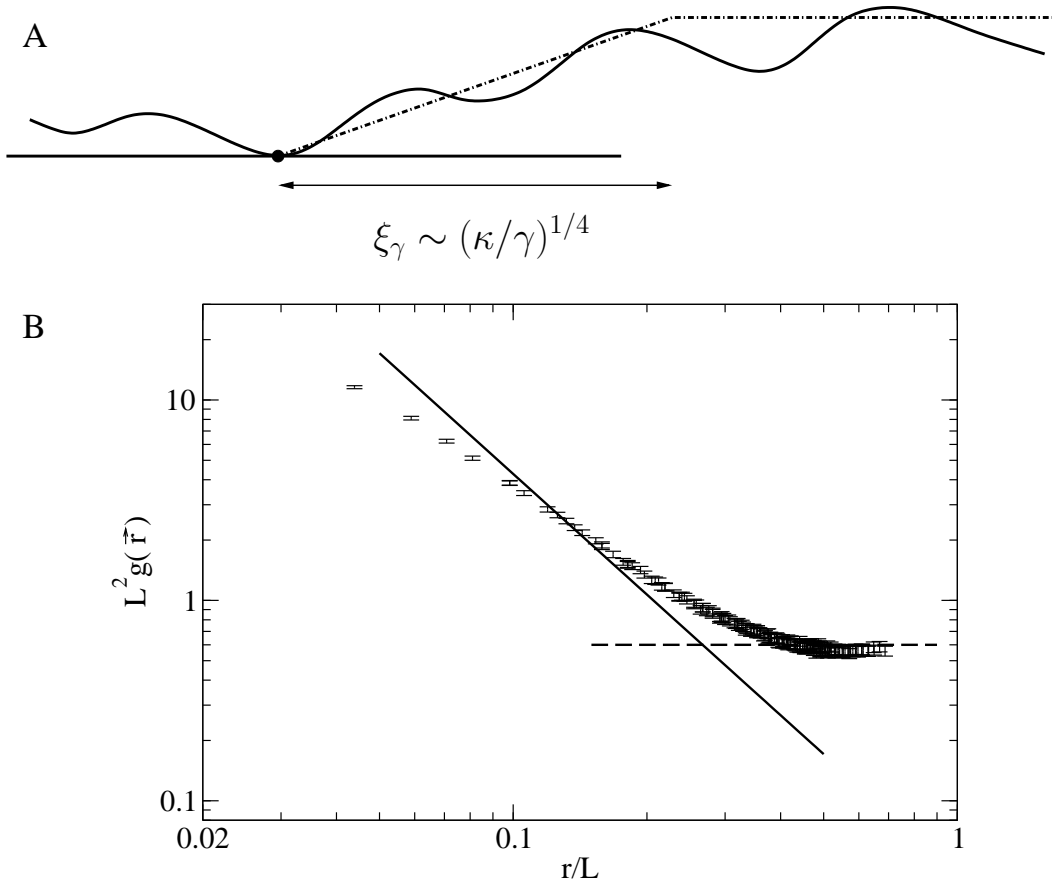,width=14cm}}
\caption{(A) The typical height of the fluctuations of a supported
membrane experiencing a harmonic confining surface potential, grows
linearly close to the pinning point and saturates at large
distances. (B) The pair correlation function, $g\left(\vec{r}\right)$,
of a such a membrane vs.~the pair distance $r$. The slopes of the
solid and dashed straight lines are $-2$ and $0$, respectively. ((B)
is adapted from {\protect \cite{farago2}})}
\label{fig:9}
\end{figure}

What if, in addition to the excluded volume repulsion, the membrane
and the surface also interact via an attractive potential of somewhat
longer range? Let us consider, for instance, the case when a harmonic
confining potential is added to the Helfrich Hamiltonian:
\begin{equation}   
{\cal H} = \int\, \left[\frac{\kappa}{2}\left(\nabla^2 h
\right)^2+\frac{\gamma}{2} h^2\right]d^2\vec{r}.
\label{eq:helfrich4}
\end{equation} 
For the harmonically confined membrane, one can define the length scale
$\xi_{\gamma}\sim (\kappa/\gamma)^{1/4}$ which marks the transition
between two scaling regimes.  For $r\ll\xi_{\gamma}$, the thermal
fluctuations are governed by the bending rigidity term in Hamiltonian
(\ref{eq:helfrich4}), while for $r\gg\xi_{\gamma}$ the harmonic
confinement term becomes dominant. The latter term is a local one,
which implies that the influence of the adhesion point becomes screened
at large distances. In the case of a single adhesion point, the height
of the fluctuations is now given by (compare with Eq.(\ref{eq:linear}))
\begin{equation}
\langle h\left(r\right)\rangle\sim\left\{
\begin{array}{ll}
r\sqrt{\frac{k_BT}{\kappa}} & {\rm for\ r\ll\xi_{\gamma}}\\
\xi_{\gamma}\sqrt{\frac{k_BT}{\kappa}} & {\rm for\ r\gg\xi_{\gamma}}
\end{array}
\right. ,
\label{eq:linear2}
\end{equation}
as illustrated schematically in Fig.~\ref{fig:9}(A).  The correlation
function of a pair of adhesion points is given by (compare with
Eq.(\ref{eq:pair1}))
\begin{equation}
g\left(\vec{r}\right)\sim\left\{
\begin{array}{ll}
r^{-2} & {\rm for\ r\ll\xi_{\gamma}}\\
r^{0} & {\rm for\ r\gg\xi_{\gamma}}
\end{array}
\right. .
\label{eq:pair2}
\end{equation} 
The results of MC simulations of an ISCG molecular model of a
harmonically confined membrane verify this crossover between the two
scaling regimes of $g(\vec{r})$ (see Fig.~\ref{fig:9}(B)).

\begin{figure}[t]
  {\centering
  \hspace{1cm}\epsfig{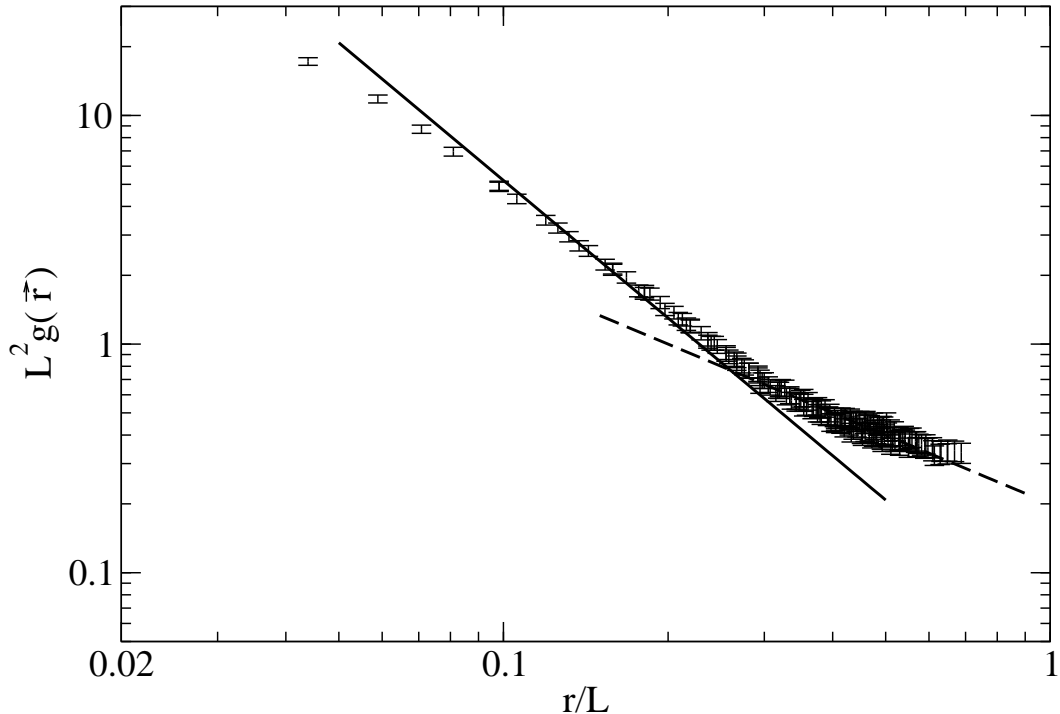}}
\caption{The pair correlation function, $g\left(\vec{r}\right)$, of a
supported membrane under tension vs.~the pair distance $r$. The slopes
of the solid and dashed straight lines are $-2$ and $-1$,
respectively. (adapted from {\protect \cite{farago2}})}
\label{fig:10}
\end{figure}

The energy of a membrane subjected to lateral surface tension
$\sigma>0$ is given by the following Hamiltonian
\begin{equation}   
{\cal H} = \int\, \left[\frac{\kappa}{2}\left(\nabla^2 h
\right)^2+\frac{\sigma}{2}\left(\vec{\nabla} h
\right)^2\right]d^2\vec{r}.
\label{eq:helfrich5}
\end{equation} 
Scaling arguments \cite{farago2} show that, in this case, the pair
correlation function exhibits behavior intermediate between
Eqs.(\ref{eq:pair1}) and (\ref{eq:pair2}): 
\begin{equation}
g\left(\vec{r}\right)\sim\left\{
\begin{array}{ll}
r^{-2} & {\rm for\ r\ll\xi_{\sigma}}\\
r^{-1} & {\rm for\ r\gg\xi_{\sigma}}
\end{array}
\right. ,
\label{eq:pair3}
\end{equation} 
where the crossover length
$\xi_{\sigma}\sim(\kappa/\sigma)^{1/2}$. This scaling form is also
confirmed by MC simulations (see Fig.~\ref{fig:10}).

\section{The strength of the fluctuation-induced attraction}

One of the main questions we aim to explore is whether the
fluctuation-induced interactions are sufficiently strong to allow the
formation of adhesion clusters. In the case of two adhesion points the
answer is no. Despite of the attractive force between the adhesion
points, they remains unbound and their mean pair separation grows
linearly with the system size $L$. More generally, if the pair
correlation function decays algebraically at large distance,
$g\left(\vec{r}\right)\sim r^{-c}$, the mean pair separation is given
by
\begin{equation}
\langle r\rangle \sim \frac{\int_l^L r^2 g(\vec{r})\,dr}{\int_l^L r 
g(\vec{r})\,dr}\sim
\left\{
\begin{array}{ll}
L & {\rm for\ c<2} \\
L/\ln L & {\rm for\ c=2} \\
L^{3-c} & {\rm for\ 2<c<3} \\
\ln L & {\rm for\ c=3} \\
l & {\rm for\ c>3} 
\end{array}
\right. .
\label{eq:pair4}
\end{equation}
The physically relevant cases in Eq.(\ref{eq:pair4}) are $c=2,1,{\rm\
and\ }0$ which, respectively, correspond to pinned, pinned-stressed,
and pinned-confined membranes. In all of these cases, $\langle r
\rangle$ grows with the size of the system. 

Another quantity of interest is the mean number $\langle C\rangle$ of
contacts between the surface and a membrane with one adhesion
point. As discussed in section \ref{sec:twopoints}, the probability
density that membrane comes into contact with the surface at a
distance $r$ from the pinning point has the same scaling form as the
pair correlation function $g\left(\vec{r}\right)$. Thus,
\begin{equation}
\langle C\rangle\sim  \int_l^L g(\vec{r})r\,dr\sim 
\left\{
\begin{array}{ll}
(L/\xi_{\gamma})^2 & {\rm for\ c=0} \\
L/\xi_{\sigma} & {\rm for\ c=1} \\
\ln(L/l) & {\rm for\ c=2} 
\end{array}
\right. .
\label{eq:contacts}
\end{equation}
We can use this last result to generalize and recalculate the
attachment free energy of one adhesion point,
Eq.(\ref{eq:attachmentfenergy}). Our original derivation of
Eq.(\ref{eq:attachmentfenergy}) was based on the argument that the
configuration phase space of a pinned membrane comprises a small
subspace within the configuration phase space of a free membrane.
More precisely, we argued that this subspace includes the free
membrane configurations in which the global minimum of the height
function occurs at the pinning point of the corresponding supported
membrane. We further argued that the relative size of the subspace is
$(l/L)^2$, which was based on the assumption that typically there is
only one contact point with the surface and, therefore, this contact
point must be the adhesion site. However, as we see from
Eq.(\ref{eq:contacts}), a typical configuration makes $\langle C\rangle$
contacts with the surface. Therefore, the partition functions of the
two problems (free vs. pinned membranes) are actually related by
$Z=\left[\langle C \rangle\left(l/L\right)^2\right]Z_{\rm free}$.  The
attachment free energy is given by
\begin{equation}
F_{\rm attachment,1}=-k_BT\ln\left(\frac{\langle C\rangle
l^2}{L^2}\right)=\left\{
\begin{array}{ll}
2k_BT\ln(\xi_{\gamma}/l) & {\rm for\ c=0} \\
k_BT\ln(L/l)+k_BT\ln(\xi_{\sigma}/l) & {\rm for\ c=1} \\
2k_BT\ln(L/l)-k_BT\ln\left[\ln(L/l)\right] & {\rm for\ c=2} 
\end{array}
\right. .
\label{eq:attachmentfenergy3}
\end{equation}
Notice that for sufficiently large $L$, $F_{\rm
attachment,1}(c=2)>F_{\rm attachment,1}(c=1)>F_{\rm
attachment,1}(c=0)$. Indeed, the attachment of a free membrane to a
surface is likely to be more costly than the attachment of stressed
and harmonically confined membranes that exhibit reduced fluctuations
and, thus, remain close to the surface anyway.

\section{The many-body problem}

Let us look back at Fig.~\ref{fig:8} which shows the pair correlation
function between two adhesion points. The figure demonstrates that the
scaling form Eq.(\ref{eq:pair1}) holds over almost the entire range of
pair separations considered ($l<r<L/\sqrt{2}$). The deviations from
the power law at small pair distances arising from the short range
depletion forces between lipids have already been discussed in section
\ref{sec:twopoints}. What is quite surprising, though, is the pretty
good agreement between the MC results and Eq.(\ref{eq:pair1}) at large
pair distances. Eq.(\ref{eq:pair1}) has been derived for two adhesion
points in a very large membrane, neglecting boundary effects. In the
simulations the conditions are different - the membrane has a finite
size and periodic boundary conditions are employed to reduce the
finite size effects. Thus, each adhesion point interacts not only with
the other adhesion point but also with its infinite array of periodic
images. Nevertheless, the existence of periodic images seems to have a
very small impact on the results. This observation is particularly
unexpected for $r>L/2$ corresponding to situations where one of the
adhesion points is equally close to two images of the other adhesion
point. The only possible way to explain this surprising observation is
to assume that the periodic images of the adhesion points are largely
screened. This assumption is consistent with the following physical
picture: The membrane mediated interactions originate from the
entropic cost due to the suppression of the membrane thermal
undulations. Thus, the presence of each adhesion point is felt only in
the region where it affects the fluctuations and cause their
reduction, while outside of this region, the adhesion point is
effectively screened.  In this perspective, the idea that distant
adhesion points are screened seems logical. The fluctuations vanish at
each adhesion point, irrespective of the distribution of the
others. Moreover, in the immediate vicinity of each adhesion point,
one expects the amplitude of the fluctuations to depend only on the
distance from that adhesion point. If the membrane is neither stressed
nor experiencing a confining surface potential, the amplitude of the
fluctuations in this region grows linearly with the distance $r$ from
the adhesion point, as given by Eq.(\ref{eq:linear}). We now wish to
introduce a more general expression that holds over the entire area of
the membrane and coincides with Eq.(\ref{eq:linear}) close to every
adhesion point. Our suggestion is as follows \cite{farago_weil}: In
each unit area of the membrane, the mean height of the membrane above
the surface is given by (compare with Eq.(\ref{eq:linear}))
\begin{equation}
\left\langle h\left(r\right)\right\rangle\sim d_{\rm
min}\sqrt{\frac{k_BT}{\kappa}},
\label{eq:linear-many}
\end{equation} 
where $d_{\rm min}$ is the distance of the unit area from the {\em
nearest}\/ adhesion point. We also replace $r$ with $d_{\rm min}$ in
Eq.(\ref{eq:fenergydensity2}) for the attachment free energy density,
which now reads
\begin{equation}
V(r)=\frac{1}{\pi}\frac{k_BT}{d_{\rm min}^2}.
\label{eq:fenergydensity-many}
\end{equation}
 The total attachment free energy of a given distribution of adhesion
 points is obtained by integrating the attachment free energy density
 Eq.(\ref{eq:fenergydensity-many}) over the entire membrane area. This
 calculation is done by constructing the Voronoi diagram of the
 distribution of adhesion points, integrating the free energy density
 with each cell (where in each cell the distance is measured from the
 adhesion point located in the cell, and a small region of microscopic
 size $l$ around the point is excluded from the integral), and summing
 the contributions of the different cells:
\begin{equation}
F_{\rm attachment}=\sum_{i=1}^{N_{\rm cell}}\int \frac{k_BT}{\pi r^2}
\,d^2\vec{r}
\label{eq:fenergy-voronoi}
\end{equation}
In a lattice-gas model, the discrete analog of this expression applies
\begin{equation}
F_{\rm attachment}
=\sum_i \frac{k_BT}{\pi}\left(\frac{l}{d_{\rm min}}\right)^2(1-s_i),
\label{eq:fenergy-lattice}
\end{equation}
where the sum run over all the empty lattice sites ($s_i=0$) and $l^2$
is the area per lattice site.

As discussed in section \ref{sec:lattice-gas}, our main goal is to
develop and use a lattice-gas model for the aggregation problem of
adhesion points. In the model, each lattice point represents an
adhesion point between the membrane and surface.  The energy of a
given configuration of lattice points is given by
Eq.(\ref{eq:latticegas2}), where the first term represents the
short-range attraction between adhesion points and the second term is
a many-body fluctuation-induced potential
$\phi\left(\left\{s_i\right\}\right)$.  Our journey to derive an
expression for $\phi\left(\left\{s_i\right\}\right)$ started in
section \ref{sec:onepoint}, and has finally reached the
end. $\phi\left(\left\{s_i\right\}\right)$ is a potential of mean
force which, for a given distribution of adhesion points, is
determined by tracing over all the relevant membrane configurations
and calculating the free energy penalty associated with the reduced
thermal fluctuations. Eq.(\ref{eq:fenergy-lattice}) provides this
expression by assigning a free energy cost with each empty lattice
site that represents a fluctuating unit area of the supported
membrane. Introducing Eq.(\ref{eq:fenergy-lattice}) into
Eq.(\ref{eq:latticegas2}), yields the energy function of our
lattice-model of adhesion points
\begin{equation}
{\cal
H}=-\epsilon\sum_{i,j}s_is_j+
\sum_i \frac{k_BT}{\pi}\left(\frac{l}{d_{\rm min}}\right)^2(1-s_i).
\label{eq:latticegasadhesion}
\end{equation} 

\subsection{The two-body problem revisited}

\begin{figure}[t]
  {\centering
  \hspace{3cm}\epsfig{file=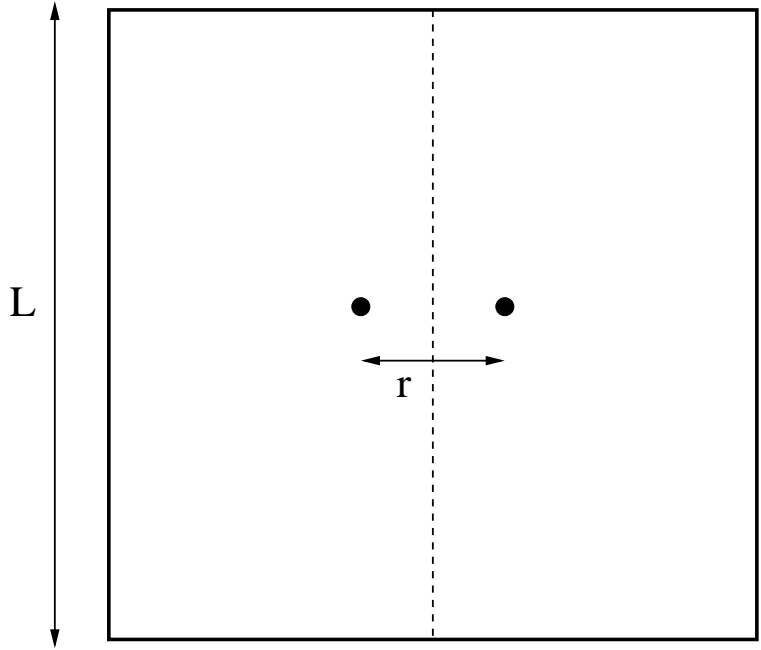,width=10cm}}
\caption{Schematic of a square membrane of linear size $L$ with two
adhesion points located at $(x,y)=(\pm r/2,0)$. The dashed line shows
the boarder between the Voronoi cells of the adhesion points.}
\label{fig:11}
\end{figure}

Let us see how one can re-derive Eq.(\ref{eq:pairpotential}) for the
pair potential of mean force by calculating the attachment free energy
Eq.(\ref{eq:fenergy-voronoi}). Towards this end, consider the membrane
shown schematically in Fig.~\ref{fig:11} with two adhesion points,
each of which located a distance $r/2$ from the center of the
membrane. The dashed line shows the boarder between the Voronoi cells
of the adhesion points, where each cell extends over half of the area
of the membrane. For the configuration shown in Fig.~\ref{fig:11}, the
attachment free energy Eq.(\ref{eq:fenergy-voronoi}) reads:
\begin{eqnarray}
F_{\rm attachment,2}&=&4\int_0^{L/2}dy\left[\int_0^{(r-l)/2}dx\,\frac{k_BT}
{\pi\left[y^2 +\left(x-r/2\right)^2\right]}\right.\nonumber \\
&+&\left.\int_{(r+l)/2}^{L/2}dx\,\frac{k_BT}{\pi\left[y^2
+\left(x-r/2\right)^2\right]}\right].
\label{eq:pairrevised1}
\end{eqnarray}
Integrating over $y$ yields, 
\begin{eqnarray}
F_{\rm attachment,2}&=&\frac{4k_BT}{\pi}\left[\int_0^{(r-l)/2}\right. 
\frac{dx}{|x-r/2|}\tan^{-1}\left(\frac{L}{2|x-r/2|}\right)
\nonumber \\
&+&\int_{(r-l)/2}^{L/2}\left. 
\frac{dx}{|x-r/2|}\tan^{-1}\left(\frac{L}{2|x-r/2|}\right)
\right].
\label{eq:pairrevised2}
\end{eqnarray}
Assuming that $l<r\ll L$, the inverse tangent function in
Eq.(\ref{eq:pairrevised2}) can be approximated by the constant value
of $\pi/2$ over most of the integration range. With this
approximation, one gets
\begin{equation}
F_{\rm attachment,2}(r,L)\simeq 2k_BT\ln\left(\frac{L}{l}\right)
+2k_BT \ln\left(\frac{r}{l}\right)
=F_{\rm attachment,1}(L)+\phi(r).
\label{eq:pairrevised3}
\end{equation}
The first term in eq.(\ref{eq:pairrevised3}) is the free energy cost
of a single adhesion site [Eq.(\ref{eq:attachmentfenergy})], which is
the expected value when the two adhesion points coincide ($r\simeq l$)
to form a single cluster. The second term, which represents the
additional free energy cost associated with the separation of the
adhesion points, is identified as the fluctuation induced pair
potential, in agreement with Eq.(\ref{eq:pairpotential}).

\subsection{Mean field theory}

We now come back to the many-body problem and start with a mean field
analysis of our lattice model Hamiltonian
(\ref{eq:latticegasadhesion}). Let us consider a lattice of $N_s$
sites of which $N\leq N_s$ sites are occupied by adhesion points. Let
us further assume that the adhesion points form $N_c\leq N$ adhesion
clusters. The free energy of system includes three contributions: (i)
the mixing entropy of the adhesion clusters, $F_{\rm mix}$, (ii) the
lattice-gas energy, $E_{\rm LG}$, of the direct interactions between
the adhesion points [first term in Eq.(\ref{eq:latticegasadhesion})],
and (iii) the attachment free energy, $F_N$ [second term in
Eq.(\ref{eq:latticegasadhesion})]. The first free energy contribution
is given by
\begin{equation}
\frac{F_{\rm
mix}}{k_BT}=N_c\left[\ln\left(\frac{N_c}{N_s}\right)-1\right]
+\frac{1}{2}c\left(\frac{N_c^2}{N_s}\right),
\end{equation}
where $c$ is the second virial coefficient. On average, each cluster
consists of $(N/N_c)$ adhesion points; and if we assume that it has a
roughly circular shape than $c\simeq 4(N/N_c)$. Denoting the number
densities of the adhesion points by $\rho=N/N_s$, and of the clusters
by $\rho^*=N_c/N_s\leq \rho$, the free energy of mixing per lattice
site is given by
\begin{equation}
\frac{F_{\rm mix}}{N_sk_BT}=\rho^*\left[\ln\left(\rho^*\right)-1\right]
+2\rho\rho^*.
\label{eq:free1}
\end{equation}

The second contribution to the free energy is due to the direct
interactions between the adhesion points.  The ground state of the
interaction energy $E_{\rm LG}$ is achieved when a single circular
adhesion domain with minimal surface is formed. If we set the ground
state as the reference energy, the energy of an ensemble of clusters
can be estimated as being proportional to the total length of the
domain boundaries. For $N_c$ circular clusters of size $(N/N_c)$ we
have
\begin{equation}
\frac{E_{\rm LG}}{N_sk_BT}=\lambda \frac{N_c}{N_s}\sqrt{\frac{N}{N_c}}
=\lambda\sqrt{\rho\rho^*},
\label{eq:free2}
\end{equation}
where $\lambda$, the associated dimensionless line tension, is
proportional to the interaction energy $\epsilon$ 
\begin{equation}
\lambda=2\sqrt{\pi}B\epsilon,
\label{eq:b_parameter}
\end{equation}
and $B$ is the mean number of nearest-neighbor vacant sites per occupied
site on the boundary of a cluster ($B\rightarrow 1$ for very large
clusters). The sum of free energy contributions (\ref{eq:free1}) and
(\ref{eq:free2}) constitutes the total free energy density (per lattice
site) of a 2D lattice gas of clusters:
\begin{equation}
\frac{F_{\rm LG}}{N_sk_BT}=\rho^*\ln(\rho^*)-\rho^*+2\rho\rho^*
+\lambda\sqrt{\rho\rho^*}.
\label{eq:freelg}
\end{equation}

The third contribution to the attachment free energy can be estimated
as follows. The clusters form $N_c$ Voronoi cells, each of which has
on average an area of $A_{\rm vor}=(N_s/N_c)l^2$. The attachment free
energy of each Voronoi cell is given by an equation similar to
Eq.(\ref{eq:attachmentfenergy}) for the attachment free energy of one
adhesion point, but with $A_{\rm vor}$ instead of the total membrane
area $L^2$. Thus
\begin{equation}
F_N=N_c\left[k_BT\ln\left(\frac{N_s}{N_c}\right)\right],
\end{equation}
and the attachment free energy density is given by
\begin{equation}
\frac{F_N}{N_sk_BT}=-\rho^*\ln(\rho^*),
\label{eq:free3}
\end{equation}
which eliminates the first term in the lattice-gas free energy density
[Eq.(\ref{eq:freelg})], yielding
\begin{equation}
\frac{F}{N_sk_BT}=\frac{F_{\rm LG}}{N_sk_BT}+\frac{F_N}{N_sk_BT}=
-\rho^*+2\rho\rho^*
+\lambda\sqrt{\rho\rho^*}.
\label{eq:free}
\end{equation}

\begin{figure}[t]
  {\centering
  \hspace{0cm}\epsfig{file=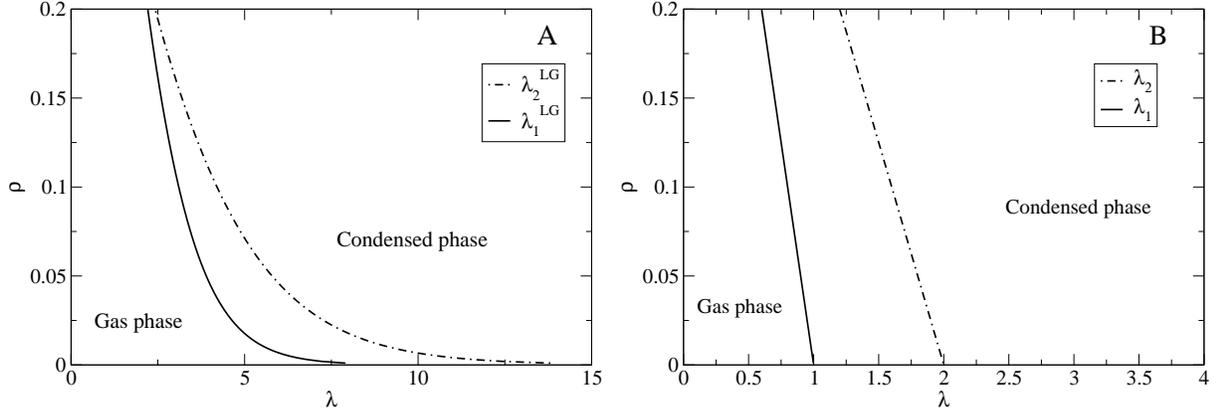,width=16cm}}
\caption{The phase diagram of the adhesion sites calculated within the
mean field approximation. (A) {\protect Eq.(\ref{eq:lambdalg})} for
the standard 2D lattice-gas model. (B) {\protect Eq.(\ref{eq:lambda})}
for adhesion points of fluctuating membranes. $\lambda_1$ and
$\lambda_2$ represent the first-order transition and spinodal lines,
respectively. (adopted from \cite{farago_weil})}
\label{fig:12}
\end{figure}

We consider a low density of adhesion sites $\rho\ll 1$, which also
implies a low number density of adhesion clusters since $\rho^*\leq
\rho$.  By minimizing the free energy density we obtain the
equilibrium value of the $\rho^*$ for the standard lattice-gas model
[Eq.(\ref{eq:freelg})] and for the adhesion points of a fluctuating
supported membrane [Eq.(\ref{eq:free})]. In both cases, the system
undergoes a first order phase transition at $\lambda_1(\rho)$ from the
gas phase ($\rho^*=\rho$) to a condensed phase consisting of only a
few clusters ($\rho^*\sim 0$). Also, in both cases, the free energy
reaches a maximum at intermediate densities ($0<\rho^*<\rho$). This
free energy barrier for condensation disappears at the spinodal point
$\lambda_2(\rho)>\lambda_1(\rho)$.  For the lattice-gas problem we
find
\begin{eqnarray}
\lambda_1^{\rm LG}&=& 1-2\rho-\ln(\rho) \nonumber \\ \lambda_2^{\rm
LG}&=& -4\rho-2\ln(\rho),
\label{eq:lambdalg}
\end{eqnarray}
while for the adhesion points of fluctuating membranes we have
\begin{eqnarray}
\lambda_1&=& 1-2\rho\nonumber \\ \lambda_2&=& 2-4\rho=2\lambda_1.
\label{eq:lambda}
\end{eqnarray}
The results of Eqs.(\ref{eq:lambdalg}) and (\ref{eq:lambda}) are
summarized in Fig.~\ref{fig:12}(A) and \ref{fig:12}(B),
respectively. The important points in the results are that: (i)
$\lambda_1>0$, which means that the fluctuation induced interactions
alone are {\em not}\/ sufficient to induce aggregation of adhesion
domains, but (ii) they greatly reduce the strength of the direct
interactions required to facilitate cluster formation since
$\lambda_1<\lambda_1^{\rm LG}$ (and also $\lambda_2<\lambda_2^{\rm
LG}$). Below, we support these conclusions with MC simulations and 
show that for adhesion points of fluctuating membranes, the site-site
cohesive energy $\epsilon$ for the onset of aggregation falls below
the thermal energy $k_BT$.

\subsection{Monte Carlo simulation}

\begin{figure}[t]
  {\centering
  \hspace{0cm}\epsfig{file=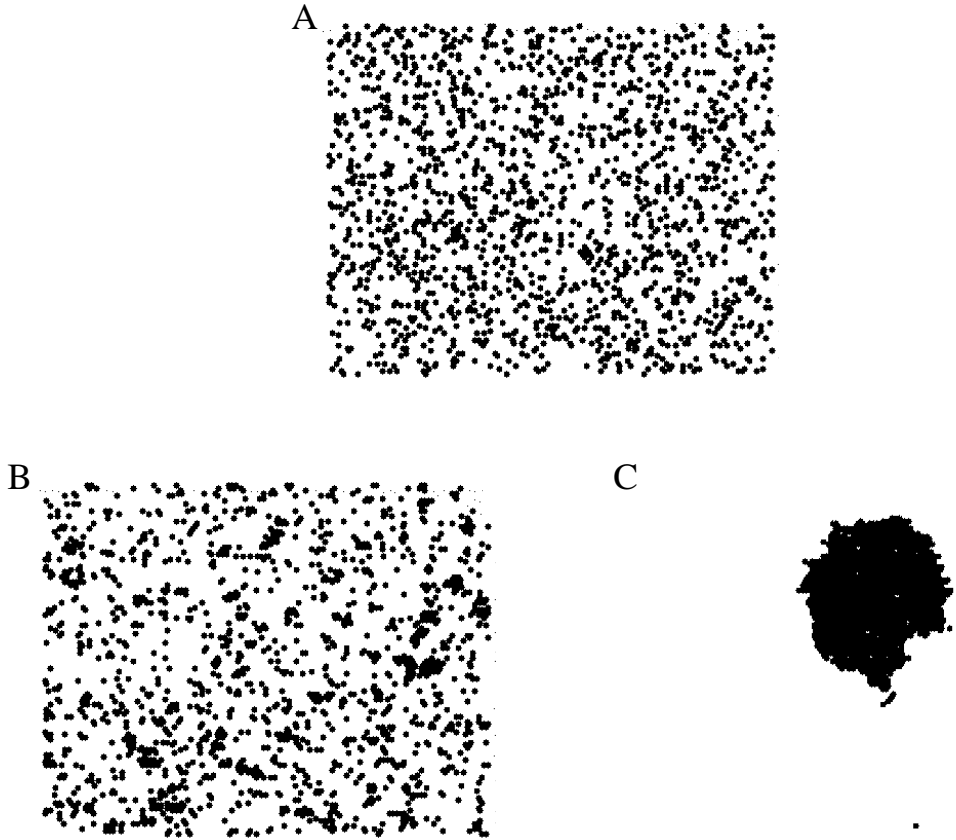,width=16cm}}
\caption{(A) Initial configurations of the simulations in which the
points are randomly distributed on the lattice. (B) Representative
equilibrium configurations of the standard lattice gas model for
$\rho=0.1$ and $\epsilon=1k_BT$. (C) Representative equilibrium
configurations of our lattice model of adhesion points for the same
values of $\rho$ and $\epsilon$ as in (B).}
\label{fig:13}
\end{figure}

To further investigate the aggregation behavior in supported
membranes, we performed MC simulations of both our lattice model of
adhesion points and of the standard 2D lattice-gas model
\cite{farago_weil}. We simulated the system at two different densities
$\rho=N/N_s=0.05$ and $\rho=0.1$, and for various values of $\epsilon$
ranging from 0 to 3 $k_BT$. Snapshots taken from simulations for
$\epsilon=1k_BT$ and $\rho=0.1$ are shown in
Fig.~\ref{fig:13}. Fig.~\ref{fig:13}(A) shows the initial
configuration where the points are randomly distributed on the
lattice.  Figs.~\ref{fig:13}(B) and \ref{fig:13}(C) show,
respectively, typical equilibrium configurations of the standard
lattice-gas model and of our model of adhesion points. One clearly
sees that for the same strength of the interaction energy
$\epsilon=1k_BT$, the standard lattice gas model remains in the gas
phase, while the adhesion points (that, in addition to the direct
interactions, also attract each other via the fluctuation-induced
mechanism) condense into a large cluster containing almost all the
adhesion points.

\begin{figure}[t]
  {\centering
  \hspace{0cm}\epsfig{file=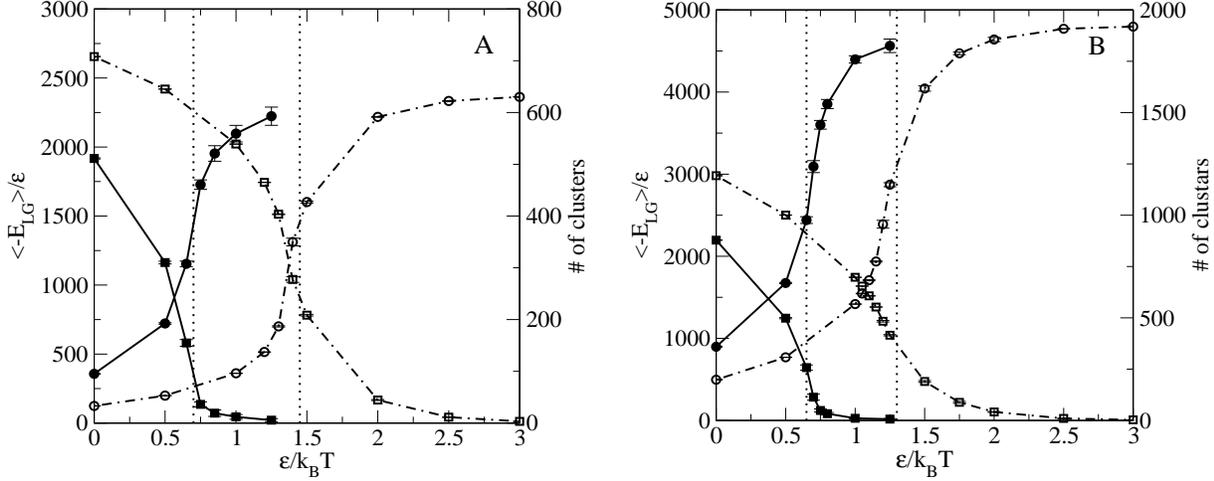,width=16cm}}
\caption{Left $y$-axis: The energy of direct interactions between
sites, $\langle E_{\rm LG}\rangle$, as a function of $\epsilon$, for
$\phi=0.05$ (A) and $\phi=0.1$ (B). Solid circles - results for our
model for adhesion points. Open circles - results for the standard
lattice-gas model. Right $y$-axis: The number of clusters as a
function of $\epsilon$, for $\phi=0.05$ (A) and $\phi=0.1$ (B). Solid
squares - results for our model for adhesion points. Open squares -
results for the standard lattice-gas model. (adapted from
\cite{farago_weil})}
\label{fig:14}      
\end{figure}

In order to determine the onset of the gas to liquid transition, we
measured the average number of clusters in the system (where a cluster
is defined as a set of neighboring occupied sites), and the mean value
of the energy of direct interactions between sites, $\langle E_{\rm
LG}\rangle$ [first term in Eq.(\ref{eq:latticegasadhesion})]. Our
results are summarized in Fig.~\ref{fig:14}(A) (for $\rho=0.05$) and
\ref{fig:14}(B) (for $\rho=0.1$). For each $\rho$, we measured these
quantities both for the standard lattice-gas model (open symbols and
dash-dotted lines in Fig.~\ref{fig:14}) and for the adhesion points
model (solid symbols and solid lines in Fig.~\ref{fig:14}). The number
of clusters is denoted by squares (values on the right $y$-axis of the
figures), while $\langle E_{\rm LG}\rangle$ is represented by circles
(values on the left $y$-axis). The gas phase is characterized by a
large number of small clusters, some of which may be of the size of a
single site. Furthermore, since each occupied site has a relatively
small number of neighboring occupied sites, the mean configurational
energy $\langle -E_{\rm LG}\rangle$ is relatively low. Conversely,
when the sites form large clusters in the condensed phase, $\langle
-E_{\rm LG}\rangle$ is high, and the total number of clusters
decreases (and in many cases, especially for large values of
$\epsilon$, we simply observe only a single cluster in our
system). Fig.~\ref{fig:14} exhibits an abrupt, clearly first-order,
transition from a gas phase with a large number of clusters and small
$\langle -E_{\rm LG}\rangle$ to a condensed state with a small number
of clusters and large $\langle -E_{\rm LG}\rangle$. The estimated
values of $\epsilon$ at the transition are (see vertical lines in
Fig.~\ref{fig:14}): $\epsilon_t\simeq0.7k_BT$ ($\rho=0.05$) and
$\epsilon_t\simeq0.65k_BT$ ($\rho=0.1$). In comparison (see also
Fig.~\ref{fig:14}), for the standard lattice-gas model, the transition
values are roughly twice larger than these values: $\epsilon_t^{\rm
LG}\simeq1.45k_BT$ ($\rho=0.05$) and $\epsilon_t^{\rm LG}\sim1.3k_BT$
($\rho=0.1$).

Our computational results which show that the fluctuation mediated
interactions reduce the strength of $\epsilon_t$, are in a qualitative
agreement with the mean field theory prediction. To make a
quantitative comparison between the theory and the simulations, one
needs to estimate the parameter $B$ appearing in
Eq.(\ref{eq:b_parameter}). Several reasons make such an estimation
difficult and inaccurate: First, our non-standard mean field theory is
based on the assumption that the clusters are circular and roughly
have the same size, which is quite a crude approximation. Second,
tracing the precise location of $\epsilon_t$ in Fig.~\ref{fig:14} is
largely inaccurate because of the finite size of the system that makes
the transitions look like crossovers. To reduce the large
uncertainties associated with the determination of $\epsilon_t$, one
can look at the difference between the value of this quantity in our
model of adhesion points and for the standard lattice-gas model. Using
\begin{equation}
\lambda_1^{\rm LG}- \lambda_1=2\sqrt{\pi}B \left(\epsilon_t^{\rm
LG}-\epsilon_t\right),
\end{equation}
for $\rho=0.1$, we find $B\simeq 1$, as indeed expected for large clusters. 

\section{Conclusions}

In this review we presented a statistical thermodynamics analysis of
the aggregation behavior of adhesion points between a fluctuating
membrane and a supporting surface. Our analysis focused on the
contribution of the membrane thermal fluctuations to this process, via
the attractive interactions that they mediate between the adhesion
points. The origin of the fluctuation-induced (Casimir-like)
interactions are the restrictions imposed on the membrane thermal
fluctuations by the adhesion points, and the associated free energy
cost which is minimized when the adhesion points localize in a
cluster. We investigated both the two- and many-body
fluctuation-induced interactions. For the two-body problem, our
analysis reveals that the fluctuation induced pair potential is
infinitely long-range with a logarithmic dependence on the pair
distance.  If, in addition to the excluded volume interactions, the
membrane and the surface also interact via an attractive confining
potential, the fluctuation-induced pair potential becomes screened at
large distances. The screening of the pair potential is due to the
fact that far away from each adhesion point, the amplitude of the
fluctuations is governed by the strength of the external potential
rather than by the presence of the other adhesion point.

In the many-body problem, the fluctuation-induced interactions are
self-screened. The amplitude of the thermal fluctuations at each unit
area of the membrane is governed by the distance to the closest
adhesion points, which implies that each point interacts only with a
few nearby points. This justifies our mapping of the problem into the
2D lattice-gas model with an effectively larger (``renormalized'')
interaction energy. Depending on the strength of the renormalized
interactions, the system may be either in a ``gas'' (uniform
distribution) or a ``condensed'' (adhesion cluster) phase. The
interesting question which arises is whether the fluctuation-induced
contribution to the attraction is sufficiently strong to allow cluster
formation. Our analysis finds that the answer to this question is
no. The fluctuation-induced interactions alone are too weak to induce
the condensation transition. They do, however, greatly reduce (to
below the thermal energy $k_BT$) the strength of the direct
interactions at which the transition takes place.

\section*{Acknowledgment}

I wish to express my deep gratitude to Prof.~Phil Pincus for inspiring
me to work on this fascinating problem. I have benefited enormously
from many insightful discussions with him. I also wish to thank my
student, Noam Weil, who participated in the final part of the research
with such enthusiasm and creativity.

The work was supported by the Israel Science Foundation (Grant Number
946/08).


\newpage

\end{document}